\documentclass[jws]{now-journal}
\usepackage[utf8]{inputenc}
\usepackage{textcomp}
\usepackage{subcaption}
\usepackage{siunitx}
\usepackage{blindtext}
\usepackage{lmodern}
\usepackage[T1]{fontenc}
\usepackage{inconsolata}
\usepackage{enumitem}
\usepackage{epstopdf}
\usepackage{algorithm}
\usepackage{algpseudocode}

\newcommand{\specialcell}[2][c]{%
	\begin{tabular}[#1]{@{}c@{}}#2\end{tabular}}

\title{Malware and graph structure of the Web}

\author[1]{Sanja \v{S}\'{c}epanovi\'{c}}
\author[2]{Igor Mishkovski}
\author[3]{Jukka Ruohonen}
\author[4]{Frederick Ayala-G\'omez}
\author[1]{Tuomas Aura} 
\author[5]{Sami Hyrynsalmi}

\affil[1]{Department of Computer Science, Aalto University, Finland; \textit{\{sanja.scepanovic,tuomas.aura\}@aalto.fi} }

\affil[2]{Faculty of Computer Science and Engineering, University Ss.\ Cyril and Methodius, Macedonia; \textit{{igor.mishkovski@finki.ukim.mk}} }  

\affil[3]{Department of Future Technologies, University of Turku, Finland; \textit{{juanruo@utu.fi}} }

\affil[4]{Faculty of Informatics, E\"{o}tv\"{o}s Lor\'and University, Hungary; \textit{{fayala@caesar.elte.hu}} }
	
\affil[5]{Pori, Tampere University of Technology, Finland; \textit{{sami.hyrynsalmi@tut.fi}} }

\creditline{Authors gratefully acknowledge the CyberTrust research project and F-Secure for their support. I.M.\ work was partially financed by the Faculty of Computer Science and Engineering at the University 'Ss.\ Cyril and Methodius'. F.Ayala-G\'{o}mez was supported by the Mexican Postgraduate Scholarship of the Mexican National Council for Science and Technology (CONACYT) and by the European Institute of Innovation and Technology (EIT) Digital Doctoral School. The authors also thank A.\ Gionis, G.\ L.\ Falher and K.\ Garimella for the helpful discussion and P.\ Hui and V. Lepp{\"a}nen for reviewing the manuscript. Special thanks to Turi for the GraphLab Academic License.}

\articledatabox{ISSN 2161-1823; DOI 10.1561/103.00000002\\
\copyright R. Banker}
\begin{document}
\begin{abstract}
Knowledge about the graph structure of the Web is important for understanding this complex socio-technical system and for devising proper policies supporting its future development. Knowledge about the differences between clean and malicious parts of the Web is important for understanding potential treats to its users and for devising protection mechanisms. In this study, we conduct data science methods on a large crawl of surface and deep Web pages with the aim to increase such knowledge. To accomplish this, we answer the following questions. 
Which theoretical distributions explain important local characteristics and network properties of websites? 
How are these characteristics and properties different between clean and malicious (malware-affected) websites? 
What is the prediction power of local characteristics and network properties to classify malware websites? 
To the best of our knowledge, this is the first large-scale study describing the differences in global properties between malicious and clean parts of the Web. In other words, our work is building on and bridging the gap between \textit{Web science} that tackles large-scale graph representations and \textit{Web cyber security} that is concerned with malicious activities on the Web. The results presented herein can also help antivirus vendors in devising approaches to improve their detection algorithms.
\end{abstract}
\section{Introduction}
Since its inception in $1989$ \citep{mcpherson2009tim} the Web has evolved and changed from a technical concept
to a complex network of networks. 
The Web is nowadays interlinking protocols, Web pages, data, people, organizations, and services \citep{hall2012web}.
Its exponential growth \citep{kleinberg1999web} results in the indexed, \textit{surface Web} with an estimated size \citep{bosch2016estimating} of $4.49$ billion pages\footnote{\url{http://www.worldwidewebsize.com/}}. At the same time, the size of the non-indexed, so-called \textit{deep web} is suggested to be $400$ to $550$ times larger \citep{bergman2001white} and rapidly expanding. In this study, we use a large Web crawl dataset provided by a security and privacy vendor F-Secure. Because of the specific crawling procedure, this dataset contains pages both from the surface and the deep Web. 

As the number of Web users increase, the number of systems that distribute malware to attack the users also increases. These attacks are done in different ways. Malware stands short for malicious software, while \textit{Web malware} is understood as malware that spreads -- and is being actively spread -- through malicious Web hosts. Web malware remains one of the most significant security threats affecting millions of hosts today, utilizing some of them for cyber crime and others as further distribution channels. 

Web science \citep{WEB-017} is a relatively new interdisciplinary field studying the complex system to which the Web has evolved. One of the challenges within Web Science is understanding the \textit{emergence phenomena} through which lower-level processes produce more complex global properties. For instance, the process of creating individual Web pages and linking them to existing ones results in a specific degree distribution of the Web graph \citep{adamic2000power}. At the same time, malicious parties might affect some of the emergence phenomena by their irregular lower-level activities. In the same example, cyber criminals have specific ways of interconnecting their malicious hosts in order to support botnets, increase possibilities of spreading malware or evade detection. This will likely yield a different degree distribution compared to creating regular websites. One of the tasks of our study is to detect and measure how malicious activities affect some of the global Web properties. 

At first, we find theoretical distributions that are the best fits to the empirical probability distributions of several website features, such as: number of pages, degree, PageRank and number of files on them. Exponentially bounded power law (truncated power law) explains well most of these distributions and each of them is a heavy tailed distribution. A result of the fitting process is also that corresponding coefficients for distributions pertaining to malicious websites differ compared to those of clean websites. In particular, we find lower power law coefficient indicating a greater skewness and irregularity of the distribution. We also study properties of a network of malicious websites based on the malware they share. Existing communities in this network reveal groups of websites devoted to sharing particular types of malware, so called malware distribution networks (MDNs). Finally, we evaluate the power of analyzed properties as predictive features for malware websites. 

On one side, we are building on Web science that studies distributions and correlations among Web properties and, on another, on Web cyber security research that characterizes and measures malicious activities. Hence our work is bridging the gap between Web science and the Web cyber security through addressing following research questions:
\begin{description}
	\item[RQ1:] Which theoretical distributions explain important local characteristics and network properties of websites in the (deep) Web?
	\item[RQ2:] How are those characteristics and network properties different between clean and malicious (malware affected) websites?
	\item[RQ3:] What is the prediction power of website local characteristics and network properties to classify malware websites?
\end{description}
The rest of the paper is organized as follows. Section \ref{sec:bckg} sets out the background and discusses related work. Section \ref{sec:data} describes Web crawling procedure, the data and our methods. Section \ref{sec:distr} present results of fitting Web distributions, where we focus on the differences between clean and malicious parts of the Web. In addition to a dichotomization to clean vs.\ malicious websites, we also analyze relative maliciousness of websites in Section \ref{sec:rel_mal_core}. Section \ref{sec:co-occur} provides insights on a malware co-occurrence network of websites.
Finally, in Section \ref{sec:prediction_power}, we discuss the prediction power of analyzed features in discerning regular from malicious websites. Discussion and conclusions are given in Section \ref{sec:discussion}.

\section{Background}
\label{sec:bckg}
A Web crawl is a dataset consisting of a set of crawled web pages, hyperlinks among them and additional metadata stored in the process. A traditional way to model this dataset as a graph has been to consider individual pages as nodes, and hyperlinks among them as directed edges. Such a representation of the Web is termed \textit{page graph}. If we consider user browsing behavior, however, a website is more a logical Web unit compared to a single page \citep{baeza2002web}. For this reason, studies have also focused on Web graph representation in which nodes represent aggregated pages from a single pay-level domain (PLD). PLD corresponds to a sub-domain of a public top-level domain (TLD), for which users usually pay for when hosting websites. Starting from uniform resource locator (URL) \citep{berners1994uniform} that uniquely identifies each page, the aggregation to PLDs is performed by extracting second level domain and TLD and concatenating them. For instance, for Aalto University's URL \texttt{http://www.aalto.fi/en/}, second level domain is \texttt{aalto}, TLD is \texttt{.fi} and PLD is \texttt{aalto.fi}. We analyze our data on such aggregation level and adopting the term used by \citet{meusel2015graph}, we operate on a \textit{PLD graph}. Technical details of the aggregation process we applied are described in Section \ref{sec:data}.

There are two main techniques for delivering Web malware to users. The first is called push-based, where the user is tricked to download the binary file using social engineering, cross-site scripting, or by related means. The second is called pull-based, where browser vulnerabilities are exploited to automatically download an exploit. The later technique is also called \textit{drive-by downloads}. In addition to pages and links among them, our dataset contains a set of binary files that are found on the pages visited. After the scanning procedure described in Section \ref{p:VirusTotal}, we assign a maliciousness score to each file. In our study, we do not distinguish between the two mentioned techniques for delivering malware (i.e., we consider all Web malware found). Using the file maliciousness information we classify PLDs as \textit{clean} (i.e., no malicious or suspicious files found on them) and \textit{malicious} (i.e., at least one such a file found). In another investigation, we also assign a relative maliciousness score to PLDs. 
\subsection{Prior work}
\textbf{Web science} \citep{berners2006creating,WEB-017} is an interdisciplinary field that emerged to tackle the Web as a complex socio-technical phenomenon. Early Web research focused on topological properties of the Web graph \citep{barabasi2000scale} and communities in it \citep{gibson1998inferring}. One of the landmark studies characterized Web structure as the famous bow-tie \citep{broder2000graph} and also suggested power law distributions for indegrees and outdegrees of pages. Interestingly, despite its importance, for a period of time after the study by \citet{broder2000graph}, other large-scale studies of the Web were rare \citep{luduena2013large} until the couple of more recent ones \citep{meusel2014graph,luduena2013large}. Suggested power law degree distributions and their inducing mechanisms were taken a matter of debate among researchers \citep{adamic2000power} and recently disproved on a larger dataset by \citet{meusel2014graph}. That study with negative result the authors performed on a \textit{page graph}, and in a follow up they analyze the same Web crawl aggregated on a PLD level \citep{meusel2015graph}. In the \textit{PLD graph}, they find a fit of indegree to power law, however for the outdegree they still conclude it is unlikely to follow a power law. In addition to distributions, correlations between important Web host features, such as indegree, outdegree and Alexa's rank\footnote{\label{note11}\url{http://www.alexa.com/topsites}} are analyzed \citep{luduena2013large}. 


Since Web use is a pervasive element of life, \textbf{Web security and privacy} became of essential importance. A number of Web security studies characterized and measured properties of malicious activities and hosts on the Web. For example, \citet{boukhtouta2015graph} used network components and their connectivity to identify malicious infrastructures. Several network node properties are employed to measure host badness, while temporal graph similarities helped to study temporal evolution of malicious infrastructures. Malicious hosts are also analyzed in terms of specificity of their life cycle properties \citep{polychronakis2008ghost}. \citet{Provos7s} performed a large scale analysis of URLs in order to describe websites performing drive-by downloads. 
\citet{invernizzi2014nazca} develop a system that detects infections by drive-by downloads in large scale networks. They analyzed Web traffic from a large Internet Service Provider. And, by considering many malware downloads together they discover malware distribution infrastructures. While some similar network analysis and machine learning methods are applied, this study is fundamentally different from ours. Firstly, the analyzed dataset represents Web traffic unlike the crawl in our study. Second, they focus on a specific type of Web malware (drive-by downloads) and do not investigate large scale Web structure and differences between clean and malware infrastructures, as is the focus of our study.



Network analysis methods have also been employed in \textbf{predicting} Web cyber-threats. For example, data about existing malware co-occurrence are used to build a file relation network and then predict new malware using label propagation \citep{ni2015file}. Another example is network analysis application for classifying malware into different families \citep{jang2014mal}. \citet{castillo2007know} showed how to successfully detect spam by analyzing the Web graph. As a next result, they also developed a classifier that combines content properties of the Web pages with link properties to successfully predict spam. 

\section{Data and methods}
\label{sec:data}
\subsection{Crawling process and statistics}
Cyber security and privacy vendor F-Secure provided the original dataset. The company collected the data from June until November 2015 using a traditional breadth-first visit crawling approach in combination with \textit{Domain Name System (DNS) brute force} crawl. The DNS brute forcing was used to expand the host data available on malicious PLDs. A large host database consisting of Alexa's $1$M top sites\footnotemark[\value{footnote}] and known link farm pages were used as the seeds for the crawler. The site scraping process used static parsing of the hypertext transfer protocol (HTML) structures. URLs from a visited site are stored in the crawl frontier and recursively visited. When a thread completed its visit to a site, it would get the next unvisited URL from the queue with prioritization policy based on website PageRank and then it would repeat the process. This procedure was continued until all URLs have been visited or a limit of $10$K outgoing links per site is reached. In addition to the pages, during the crawl, all \textit{binary files} with extensions \texttt{exe, swf, jar, zip, tar.gz} under $5$MB in size were downloaded. During the crawling period, around $120$ billion unique links are discovered and $2.5$ billion pages are visited, resulting in around $95$ terabytes of HTML content stored. The number of unique links leading to a binary download was $2.9$ million, resulting in more than $1.6$ million unique binary files stored (see Table \ref{t:data}).
\begin{table}
	\centering
	\caption{Dataset statistics}\label{t:data}
	\begin{tabular}{lc} \hline
		\textbf{\texttt{crawl element}}&\textbf{\texttt{totals}}\\ \hline
		pages & $\sim 2.5 \cdot 10^9$ \\ 
		PLDs & $6\,523\,861$ \\ 
		unique links to files & 2\,850\,868\\ 
		binary files & 1\,639\,708\\ 
		PLDs with files &221\,305\\ \hline
	\end{tabular}
\end{table}

At this point, it is imperative to remind that different crawling policies and limits imposed are likely to affect the resulting dataset and have potential to induce certain biases. As the focus in this work is to unveil in particular the malware distributions and properties of malicious PLDs on the Web, using Alexa's top sites as part of the seed might seem as a suboptimal choice. However, there are several benefits to using such a seed, as we detail in the following. First, larger seed sets are known to make the crawl more stable. Moreover, seeding our crawl with malicious hosts from the start would not be optimal for intended host discovery. From a technical viewpoint, Alexa's top sites listing contains the information that were required by our crawling procedure, such as about PageRank. Second, despite the focus on analyzing malware, we also want to investigate its position in the \textit{regular Web}, as typical users might experience it. Starting, from the most popular sites on the Web can give us a picture of \textit{how many clicks away} a typical user is from accessing malware sites. Finally, we also compare malware files and PLDs to the clean PLDs and non-malicious files, and so we need a good and representative coverage of the normal, clean, portion of the Web. 

\subsection{PLD graph}
As mentioned in the introduction, a PLD graph is created from the page graph by aggregation on the PLD level.
In order to aggregate page URLs to their corresponding PLDs, we use library \texttt{TLDextract} \citep{TLDextract} that is looking up the Mozilla's initiative Public Suffix List\footnote{\url{https://publicsuffix.org/}} for most up to date TLDs. Throughout the rest of the paper we use the term \textit{PLD} as a synonym for a 2-LD + 1-LD\footnote{1-LD can be generic top-level domain (gTLD) or country code top-level domain (ccTLD) } in $\lambda$-notation introduced by \citet{berger2016mining}, where $\lambda$-LD is $\lambda$-level domain. As an example, if we have nodes \texttt{a.2.com.cn/index.html} and \texttt{b.2.com.cn/foo/bar/baz.html} in the \textit{page graph}, then in the \textit{PLD graph}, we aggregate them to a single node \texttt{2.com.cn}. Note that in this example, 1-LD is \texttt{.com.cn}, and not \texttt{.cn}. Similarly, if we have a link from \texttt{a.2.com.cn} toward \texttt{b.2.com.cn} in the \textit{page graph}, then in the \textit{PLD graph} we have a self-loop at \texttt{2.com.cn}.

The resulting \textit{PLD graph} $G=(V, E)$, that we focus on, has $|V|=6\,523\,861$ distinct nodes connected by $|E| = 111\,273\,135$ edges with an average node degree of $47.2$, seven times higher compared to the page graph. The distribution of the number of aggregated pages per PLD is shown in Fig.\ \ref{fig:pages_per_domain}. There is a small number of domains that have even more than $100\,000$ pages; \texttt{blogspot}-domains are prominent in this top list. Most of the domains host only $1$ to $3$ pages.

\subsection{PLD file diversity} 
As a measure of file diversity on a single PLD, we apply information entropy measure \citep{shannon2001mathematical}. For a PLD, we denote the number of unique files present on it as $N$, and the total number including file copies as ${TF}$. For each unique file $f_i$ found $k_i$ times on the PLD, we assign the file probability $p_{f_i} = {k_i}/{TF}$. Now we have the file distribution probabilities $P = p_{1}, .., p_{N}$ for each domain and we calculate the entropy:
\begin{equation}
H = - \sum_{i=1}^{N} {p_{i} \cdot \log_2{p_{i}}}.
\end{equation}
PLDs with more unique files, in general, will have higher $H$ values, while less file diversity (or more copies) will lead to a decreased $H$. 

\subsection{PLD malware co-occurrence subgraph} In order to further characterize the malicious PLDs, we build another type of a network -- based on the shared malware files. In the \textit{domain malware co-occurrence network} $M = (V_m, E_m, w)$, the node set $V_m$ consists of PLDs on which malicious files are found. The undirected edges set $E_m$ contains the links between the PLDs that have hosted at least one common malicious file. The weight $w \in (0,1]$ for an edge is defined as Jaccard similarity of the sets of malware files hosted on the two PLDs connected by the edge.

\subsection{File reputation} 
\label{p:VirusTotal}
We enrich the Web crawl by scanning each of the $\sim 1,6$M file hashes through VirusTotal API\footnote{\url{www.virustotal.com}}. An independent maliciousness score is given to the file scanned by each of $d=56$ antivirus (AV) engines that are included in the VirusTotal service. We take a similar approach as in our previous study \citep{ruohonen2016post} and calculate overall maliciousness score $\bar{\delta}$ of a file ${f_i}$ using the formula:
\begin{equation} \label{eq:1}
\bar{\delta}({f_i}) = s\bigg(\frac{1}{d} \sum_{k=1}^d \delta_{k} ({f_i}) \bigg), \text{   } s(x) =
\begin{cases}
0 & \text{if } x \leq \tau, \\    
1 & \text{otherwise};
\end{cases}
\end{equation}
where $\delta_{k}({f_i}) \in \{0,1\}$ is the score given to the file by the $k$-th AV engine. Selecting the threshold $\tau \in [0, 1)$ within $s(x)$ is used to dichotomize the score depending on how strictly we want to define malware or whether we want to focus also on suspicious and potentially unwanted files. 
An earlier study \citep{lindorfer2014andrubis} found that if only $5$ of the VirusTotal engines have marked the file as malware, it can be considered malicious, while \citet{invernizzi2014nazca} used threshold of $2$ as a proxy for maliciousness. At the same time, the analysis of AV detection rates revealed that the best engine had an average detection of $70\%$ \citep{Provos7s}. Moreover, there is a time lag until AV engine virus definitions are updated, and if one scans suspicious files at a later time (2 months in the case of study ibid.), AV engines will flag more suspicious files as malicious. Considering such results and the statistical trade-off that a stricter threshold $\tau$ reduces the size of our malware set, in the first part of the analysis we set $\tau=0$, i.e., to the highest alert level. After such an evaluation procedure, our malware set consists of $45\,172$ files. 

In the second part of this study, we also consider a ratio-based maliciousness score $\bar{\rho}$ defined as:
\begin{equation} \label{eq:2}
\bar{\rho}({f_i}) = \frac{1}{d} \sum_{k=1}^d \delta_{k} ({f_i}).
\end{equation}

\subsection{PLD reputation} 
\label{p:dom_mal_score}
Based on the type of files that populate them, PLDs are, in similarity to files, given two types of maliciousness scores. For the first part of this study, introducing dichotomous maliciousness score, we categorize PLDs using following (strict) procedure. A PLD is considered \textit{clean PLD} if no malware files, as defined by Eq.\ \ref{eq:1}, are found on it. PLDs having at least one malicious file are considered \textit{malicious PLDs}. Using such a dichotomization, among $221\,305$ PLDs hosting at least one file of any type, we find $11\,242$ malicious PLDs ($\sim$5\%).

In the second part of the study, we assign following ratio-based score to PLDs:
\begin{equation} \label{eq:3}
\bar{r}({{PLD}_i}) = \frac{1}{{TF}_{PLD}} \sum_{i=1}^{{TF}_{PLD}} \bar{\rho}({f_i}),
\end{equation}
where ${TF}_{PLD}$ is the total number of files (including clean) found on the PLD. Introduced score $\bar{r}({{PLD}_i})$ will equal to $0$ for all clean PLDs from the dichotomization above, while the malicious PLDs will receive a score $0 < r \leq 1$ quantifying the share of malicious files to all the files, and also the maliciousness of those files, as per Eq.\ \ref{eq:2}.  

\subsection{Domain name entropy} \label{sec:PLD_name}
Domain Generation Algorithms (DGAs) yield a large number of pseudorandom domain names generated using a seed value precalculated by the attackers. DGAs have malicious applications for dynamical provision of command and control centers (C\&C), drive-by download attacks and spam domains creation \citep{sood2016taxonomy}, among others. DGAs are likely to result in domain names that follow some pattern of creation, in contrast to real words that are most often used by humans in regular domain names \citep{yadav2010detecting}. For example, algorithmically generated domain names might have following format \texttt{cxxx.com.cn}, where $x \in$ \texttt{a...z} (example from our dataset).  

Several more or less sophisticated approaches are proposed for detecting such algorithmically generated domain names \citep{yadav2010detecting,demertzis2015evolving}. For our purpose, we find that relatively simple \textbf{domain name badness score} \citep{SANSfreq.py} is sufficient. We will also in short refer to this score as \textbf{domain name entropy}. The score is based on a \textit{frequency table} of adjacent character pairs within regular English text. For instance, normal English text is likely to feature character pairs such as \textit{th}, \textit{qu} or \textit{er}, but unlikely to feature \textit{wz} or \textit{dt}. The expected frequencies of regular names are calculated from Alexa's top $1$M most common website names and also texts from the literature. Once the frequency table is built, far a given domain name we lookup the table for frequencies of character pairs within the name and estimate how probable it is to represent a regular domain name. This approach is shown to differentiate well normal domain names from algorithmically generated ones. Former can be characterized with the badness score higher than $5$ and later with the score lower than $5$ \citep{SANSfreq.py}. We implement such a score and use it to assess some of the irregularities in our data.

\subsection{Fitting heavy tailed distributions}
\label{sec:fitting}
\begin{table} 
	\centering
	\caption{\textbf{Heavy tailed distributions assessed in our fitting procedure.} Table adapted from \citep{clauset2009power}. Probability $p(x) = C f(x)$, for some constant $C$. }\label{t:heavy-tailed_distr}
	\begin{tabular}{lll} \hline
		\textbf{\texttt{distribution}}& \texttt{\texttt{{$f(x)$}}} & \textbf{\texttt{parameters}} \\ \hline 
		power law & $x^{-\alpha}$ & $\alpha$ \\ 
		truncated power law &  $x^{-\alpha} e^{-\lambda x}$ & $\alpha, \lambda$\\ 
		exponential & $e^{-\lambda x}$ & $\lambda$ \\ \hline
		stretched exponential & $x ^{\beta-1}e^{-\lambda x^{\beta}}$ & $ \beta, \lambda $ \\ 
		log-normal & $ \frac{1}{x} e^{ {- \frac{(ln x - \mu )^2}{2 {\sigma}^2}} } $ & $\mu, \sigma$ \\ 
		log-normal positive & $ \frac{1}{x} e^{ {- \frac{(ln x - \mu )^2}{2 {\sigma}^2}} } $ & $\mu > 0, \sigma $ \\ \hline
	\end{tabular}
\end{table}
The first part of this study is concerned with fitting distributions of the website features, many of which are suggested to be heavy tailed \citep{clauset2009power,broder2000graph,meusel2014graph,meusel2015graph}. In this subsection we describe the methods and tools that are used in our fitting procedure. Theoretical foundations about power law distributions in empirical data are established in their seminal paper by \citet{clauset2009power}. Since power law distributions are considered among the most interesting observations in many disciplines, including physics, computer science, economics, political science and psychology, Clauset et al.\ have presented a model for fitting \textit{power law} to empirical data. Their model is, however, easily applicable to other types of theoretical distributions. We use their model to assess several heavy tailed distributions listed in Table \ref{t:heavy-tailed_distr} as plausible hypotheses to explain our Web distributions. The tool we employ is Python \texttt{powerlaw} package by \citet{alstott2014powerlaw} that implements the fitting procedure for all those distributions. 

Usually empirical data will follow a heavy tailed distribution only for some part of the data, i.e., for the values larger than some lower bound $x_{min}$ (the tail). \citet{clauset2009power} describe three main steps in their framework for analyzing power law distributed data. Below we summarize these steps as they would apply to any heavy tailed distribution $f(x)$:
\begin{enumerate}
	\item estimate $x_{min}$ and the respective parameters (Table \ref{t:heavy-tailed_distr}) of the $f(x)$ model,
	\item calculate the goodness-of-fit between the empirical data and the estimated model, 
	\item compare $f(x)$ against other plausible hypotheses via likelihood ratio test.
\end{enumerate}
In \textbf{step 1.}, $x_{min}$ is estimated using Kolmogorov-Smirnov (KS) statistics \citep{seiler1989numerical}. Such $x_{min}$ is selected for which the probability distributions of the hypothesized model and empirical data are most similar \citep{clauset2007frequency}. If one would select too low $x_{min}$ then the KS statistics would show a larger difference since we would be trying to fit a heavy tailed $f(x)$ to a part of the data that is not heavy tailed. On the contrary, a too high $x_{min}$ would result in throwing away a large part of the data that are actually well explained by the heavy tailed $f(x)$. This would in turn increase the bias from finite size effects and make the KS statistics between the distributions higher due to statistical fluctuations. After establishing the $x_{min}$, parameters of $f(x)$ are selected using maximum likelihood estimators (MLEs) \citep{cox1994inference, wasserman2013all}. 

In \textbf{step 2.}, a goodness-of-fit test should answer to the whether hypothesized $f(x)$ is a plausible explanation for the given data. Goodness-of-fit test is in this case applied as follows. One estimates the difference between the empirical and hypothesized theoretical distributions using KS statistics. Afterwards, comparable estimates are made for a number of synthetic datasets drawn from the hypothesized model. If the estimated difference for the empirical data is not importantly larger than for synthetic data, then $f(x)$ is a plausible fit to the data.

In \textbf{step 3.}, log likelihood ratio is calculated between hypothesized $f(x)$ and competing distributions plausibly explaining the data. In our case, we always compare against all the other heavy tailed distributions presented in Table \ref{t:heavy-tailed_distr}. The sign of the log of the ratio of the two likelihoods $\mathcal{R}$ tells which distribution is a better fit and $p$-value is calculated for significance of the result (for details see Section 5.1.\ in \citep{clauset2009power}). 

\texttt{Powerlaw} package implements steps 1.\ and 3., but not step 2. One reason is that step 2.\ is not necessary in those cases when it turns out in step 3.\ that some other distribution is a better fit to the data. Moreover, the presented goodness-of-fit test in step 2.\ is often too strict for any empirical dataset of a large enough size having some noise or imperfections to pass it \citep{alstott2014powerlaw, klaus2011statistical}. Hence, if one is not concerned with whether their data strictly follow a certain theoretical distribution, but instead which distribution is the best description available, then steps 1.\ and 3.\ are enough and those are the steps we apply in our fitting procedure. 

The whole \textbf{fitting procedure} that we apply can be summarized now as follows: 
\begin{enumerate}
	\item Start with an empty set of candidate distributions $\mathcal{C} = \emptyset$. 
	\item Consider each heavy tailed distribution $f(x)$ from Table \ref{t:heavy-tailed_distr} as a candidate distribution (\texttt{candidate = True}) and:
	\begin{enumerate}[label*=\arabic*.]
		\item estimate $x_{min}$ and respective parameters of the $f(x)$ model, 
		\item compare $f(x)$ against other distributions $g(x)$ from Table \ref{t:heavy-tailed_distr} via likelihood ratio test. If resulting $\mathcal{R}<0$ and $p<0.01$, then $f(x)$ is not anymore a candidate and return \texttt{False}.
		\item if the subporcedure from the previous step returned \texttt{True}, then $\mathcal{C} = \mathcal{C} \cup f(x)$
	\end{enumerate}
	\item If $|\mathcal{C}| = 1$ then a single best fit distribution is found; otherwise, evaluate the set of candidates additionally using human judgment. In this step, we take into account the concrete empirical distribution we are fitting and the mechanisms of its real-world creation to help us in deciding among the set of found candidate distributions. A fitting distribution selected in this way is marked with $*$ to distinguish it from the cases when $|\mathcal{C}| = 1$, i.e., one distribution was strongly preferred over all the others.
\end{enumerate}
For example, if the returned set of candidates $\mathcal{C}$ consist of two distributions: power law and log-normal, we might select the fit as follows. Log-normal distribution can be created by \textit{multiplying} random variables (since the log of log-normal distribution is a normal distribution that can be created by \textit{adding} random variables). If the parameter $\mu$ for log-normal fit is negative (i.e., it is not a log-normal positive distribution), then this would require such random variables to be typically negative. The Web distributions we evaluate, such as number of pages, degree, PageRank and total number of files on a PLD, are unlikely to be generated by a process that multiplies negative values. So in this case, we would select the power law distribution fit.
As another example, say we evaluate, for instance, the number of pages, on the whole PLD graph $G$ and find that $\mathcal{C}_{G} = \{f(x)\}$. In the future investigation we might evaluate the same feature (number of pages) on a subset of PLDs containing files $G_f$, which will be importantly smaller in size. If in this case we find $\mathcal{C}_{G_{f}} = \{f(x), g(x)\}$, then we will select the fit to be $f(x)$, as it is likely that a subset of a larger dataset follows the same distribution, but due to statistical fluctuations of a smaller sample set, we did not find $f(x)$ strongly preferred over $g(x)$.

\subsection{Prediction methods}
In our experiments we use well known classification methods such as Support Vector Machines (SVM) \citep{cortes1995support}, Gradient Boosting Trees \citep{friedman2002stochastic} and Logistic Regression. We also included preprocessing steps such as Synthetic Minority Over-sampling Technique (SMOTE) \citep{chawla2002smote}, cluster-based \citep{zhang2010cluster} and random majority undersampling with replacement \citep{imbalanced-learn}.
\section{Distributions of PLD features}
\label{sec:distr}

\subsection{Distributions of local PLD characteristics}
\begin{table} 
	\centering
	\caption{Percent of TLDs for the PLDs in the distribution peaks in Fig.\ \ref{fig:pages_per_domain} }\label{t:pages_per_domain}
	\begin{tabular}{cllll} \hline
		\textbf{\texttt{Range / gTLD}}&\texttt{.com}&\texttt{.pw}&\texttt{.cn}&\texttt{.xyz}\\ \hline 
		$(150,200)$ &$0.79$&$\textbf{0.07}$&$0.01$&<$0.01$ \\ 
		$(300,400)$ &$0.79$& <$0.01$& $0.05$& $\textbf{0.12}$\\ 
		$(600,700)$ &$0.83$& <$0.01$& $\textbf{0.09}$&<$0.01$ \\ 
		>$700$ & $0.20$& $\textbf{0.40}$& $0.17$&<$0.01$ \\ \hline
	\end{tabular}
\end{table}

\begin{figure}[!ht]
	\begin{subfigure}{0.97\linewidth}
		\centering
		\includegraphics[width=0.95\linewidth]{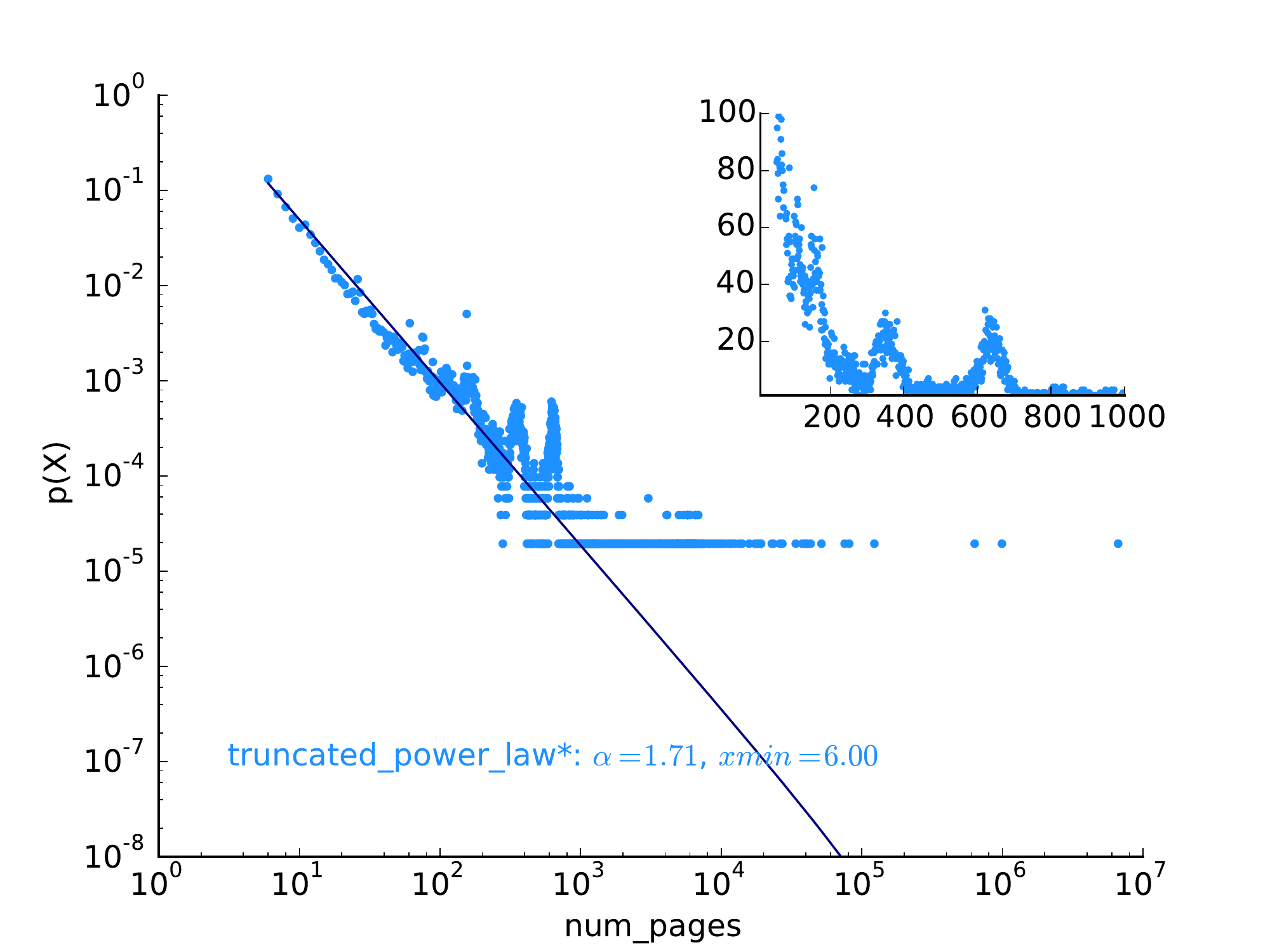} 
		\caption{\textbf{whole PLD graph.} In the inset we zoom in to the three peaks in the distribution: from around $150$ to $200$, $300$ to $400$ and $600$ to $700$ pages per domain (inset axes in linear scale).}
		\label{fig:pages_per_domain}
	\end{subfigure}  \newline
	\begin{subfigure}{0.97\linewidth}
		\centering
		\includegraphics[width=0.95\linewidth]{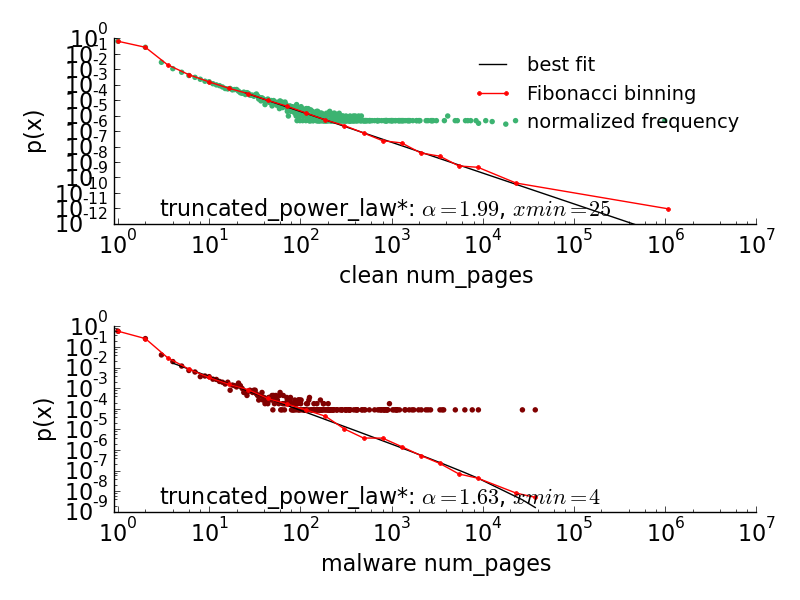}
		\caption{\textbf{clean vs. malicious PLDs.} Fibonacci binning is used to visualize the empirical distributions \citep{vigna2013fibonacci}. }
		\label{fig:pages_per_domain_CM}
	\end{subfigure}
	\caption{\textbf{Distribution fits for the number of pages}}
\end{figure}

\begin{table} 
	\centering
	\caption{\textbf{Fitting local PLD characteristics distributions.} m denotes malicious PLD subset and c the clean.\label{t:fit-loc}}
	\begin{tabular}{lll} \hline
		\textbf{\texttt{property}} &  \textbf{\texttt{best fit}} & \texttt{\textbf{parameters}} \\ \hline
		\# pages & trunc.\ power law &  $\alpha=1.71, \lambda=6.60e^{-6}$ \\ 
        \# pages m & trunc.\ power law* &  $\alpha_m=1.66, \lambda_m=3.19e^{-05}$  \\ 
        \# pages c & trunc.\ power law* &  $\alpha_c=1.99, \lambda_c=2.18e^{-11}$ \\  \hline
        tot.\ files m & lognormal pos.* &  $\mu_m=7.28e^{-5}, \sigma_m=2.77$  \\ 
        tot.\ files c & lognormal &  $\mu_c=-27.19, \sigma_c=5.47$  \\  \hline
        uniq.\ files m & trunc.\ power law* &  $\alpha_m=1.51, \lambda_m=7.98e^{-5}$  \\  \hline
        uniq.\ files c & trunc.\ power law &  $\alpha_c=2.07, \lambda_c=7.67e^{-5}$  \\ 
	\end{tabular}
\end{table}
The \textit{number of pages distribution} in PLD graph $G$ is visualized in Fig.\ \ref{fig:pages_per_domain}. We present also its best fit -- to the \textit{power law with exponential cutoff} starting from $x_{min}=4$ (for the details of the fitting procedure, see Data and methods \ref{sec:fitting}). The inset offers a first example of how malicious activity affects Web distributions. Namely, the irregularity of the distribution in the three peaks suggest possible malicious activity. Indeed, we find a larger percent ($58\%$) of PLDs having domain name badness score lower than $5$ in the three peaks compared to the rest of the distribution ($19\%$). As discussed in Section \ref{sec:PLD_name}, domain name badness score lower than $5$ is indicative of DGA activity. We also detect a larger percent of domains having such a low domain badness score in the distribution's long tail (the PLDs with more than $700$ pages). Results of inquiry into the PLDs causing the peaks are summarized in Table \ref{t:pages_per_domain}. While \texttt{.com} TLDs are most common throughout most of the distribution, the tail is dominated by \texttt{.pw} TLDs. Other TLDs found common in the peaks are \texttt{.xyz} and \texttt{.cn} (ccTLD for China). As a remark, both \texttt{.xyz} and \texttt{.pw} are relatively newly available TLDs to the general public and they are used by legitimate registrants. However, a sudden increase in the number of new registrants for both TLDs in recent years\footnote{\url{http://www.thedomains.com/2016/01/10/xyz-blows-past-us-which-had-a-28-year-head-start/}} is in agreement with our results connecting them to potential malicious activity. Symantec, for instance, released reports about the rise of spam messages from \texttt{.pw} domains\footnote{\url{http://www.symantec.com/connect/blogs/pw-urls-spam-keep-showing}}. As for the ccTLD of China, in the following parts of the study we confirm that indeed the largest percent of malicious PLDs in our dataset have that TLD (similar result reported in \citep{Provos7s}).

In Fig.\ \ref{fig:pages_per_domain_CM} we present the same distribution dichotomized by the PLD maliciousness score, separate for clean and malicious PLDs. There are a couple of interesting results from this analysis: the distributions follow a different power law coefficient $\alpha$ for the two classes of domain, and $\alpha_m$ of the malicious class is lower compared to $\alpha_c$ of the clean class. Lower $\alpha$ means higher skewness of the distribution, and is intuitively in agreement with malicious PLDs exhibiting higher irregularity in their properties. It is also interesting the $\alpha_c >2$ and $\alpha_m<2$, as it means that the two power law-like distributions qualitatively differ. For instance, it means that the clean distribution has a well defined mean, while malicious does not \citep{newman2005power}. 
Table \ref{t:fit-loc} summarizes the fitting results for other local PLD characteristics evaluated. For the total number of files lognormal and lognormal positive are found the best fits. In the case of the number of unique files, we find exponentially bounded power law to explain best the distribution, and again it holds: $\alpha_m < \alpha_c$.

\begin{table} 
	\centering
	\caption{\textbf{Fitting network PLD properties distributions.} m denotes malicious PLD subset and c the clean. PR stands short for PageRank and tc for triangle count.\label{t:fit-netw}}
	\begin{tabular}{lll} \hline
		\textbf{\texttt{property}} &  \textbf{\texttt{best fit}} & \texttt{\textbf{parameters}} \\ \hline 
		indeg & trunc.\ power law &  $\alpha=1.66, \lambda=2.43e^{-4}$ \\ 
        indeg m & trunc.\ power law* &  $\alpha_m=1.61, \lambda_m=4.62e^{-6}$  \\ 
        indeg c & trunc.\ power law* &  $\alpha_c=2.21, \lambda_c=5.96e^{-12}$ \\  \hline
		outdeg & trunc.\ power law & $\alpha=1.70, \lambda=2.01e^{-4}$\\
        outdeg m & trunc.\ power law* & $\alpha_m=1.97, \lambda_m=8.61e^{-8
}$\\
        outdeg c & trunc.\ power law* & $\alpha_c=2.06, \lambda_c=8.76e^{-8}$\\ \hline
        PR m & trunc.\ power law* & $\alpha_m=1.61, \lambda_m=1.97e^{-5}$\\
        PR c & trunc.\ power law* & $\alpha_c=1.93, \lambda_c=3.07e^{-5}$\\ \hline
        tc m & stretched exp* & $\beta=0.76, \lambda_m=7.24e^{4}$\\
        tc c & lognormal positive* & $\mu_c=3.88, \sigma_c=1.18e^{-6}$\\ \hline
	\end{tabular}
\end{table}
\begin{figure*}[!ht]
	\centering
	\begin{subfigure}{.49\linewidth}
		\centering
		\includegraphics[width=\linewidth]{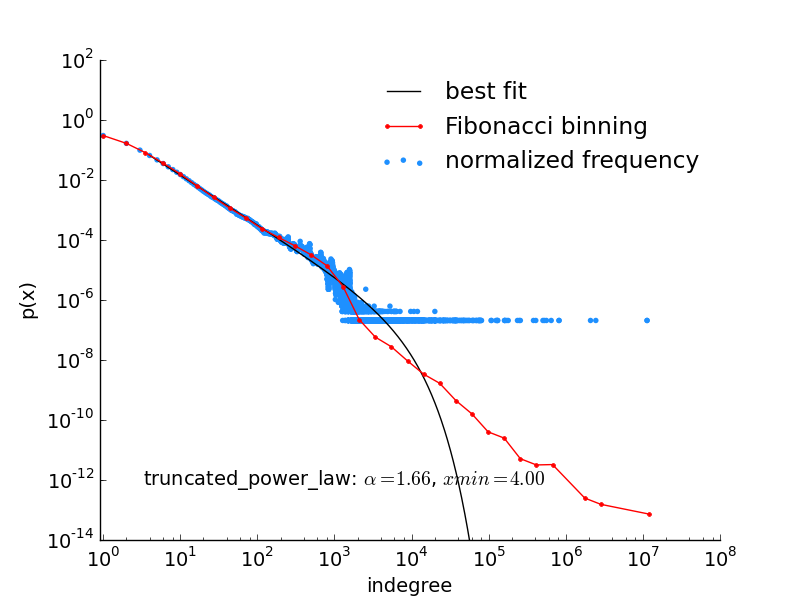}
	\end{subfigure}
	\begin{subfigure}{.49\linewidth}
		\centering
		\includegraphics[width=\linewidth]{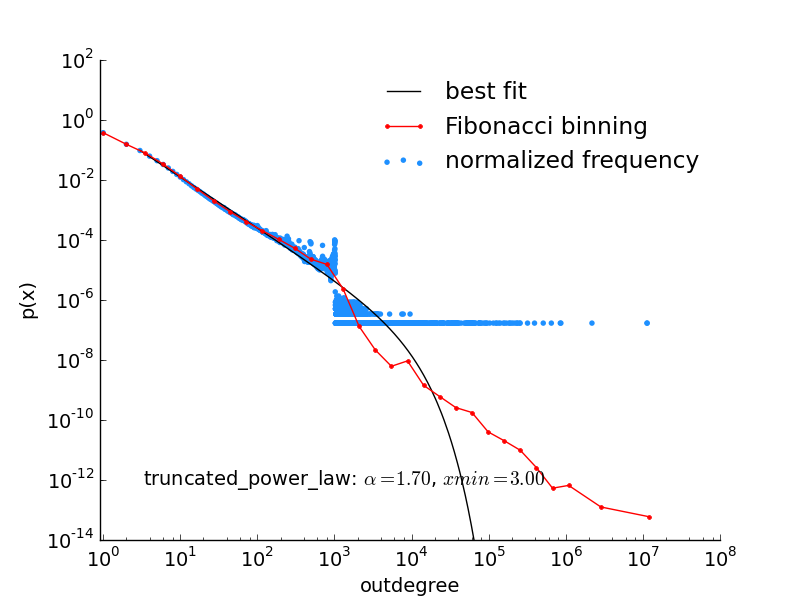}
	\end{subfigure}
	\begin{subfigure}{.49\linewidth}
		\centering
		\includegraphics[width=\linewidth]{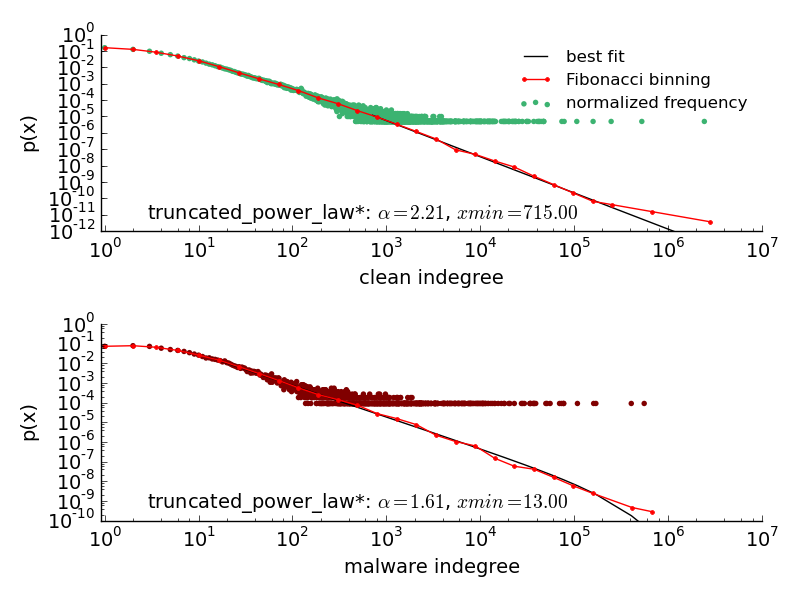}
	\end{subfigure}
	\begin{subfigure}{.49\linewidth}
		\centering
		\includegraphics[width=\linewidth]{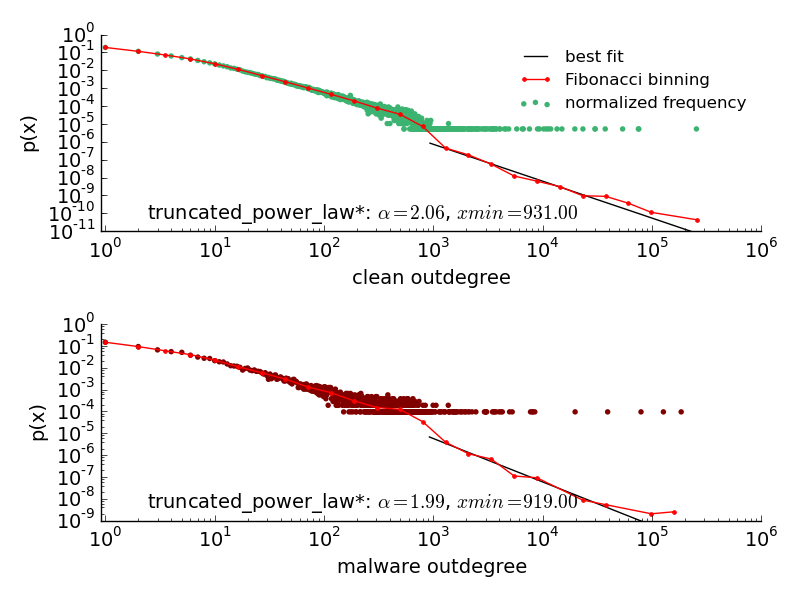}
	\end{subfigure}
	\caption{\textbf{Distribution fits for indegree and outdegree:} the whole PLD graph (top) and clean and malicious PLDs (bottom). Fibonacci binning is used to visualize the empirical distributions \citep{vigna2013fibonacci}.}
	\label{fig:degree}
\end{figure*}
\subsection{Distributions of network PLD properties} 
Network properties that we investigate in PLD graph are degree, PageRank \citep{brin2012reprint}, hubs and authorities scores using HITS algorithm \citep{kleinberg1999authoritative}, and number of triangles. We present the fitting results only for degree distributions, while for others we summarize the results.

Indegree and outdegree distributions for the whole $G$ and separated between clean and malicious PLDs are presented in Fig.\ \ref{fig:degree}. As mentioned in Background \ref{sec:bckg}, in their analysis of a \textit{PLD graph}, \citet{meusel2015graph} find a fit to power law for indegree distribution from $x_{min}=3\, 062$ and for outdegree they suggest that it is unlikely to follow a power law. In our dataset, we find that both empirical distributions are best explained by exponentially bounded power law (truncated power law) (see Data and methods \ref{sec:fitting} for details of the fitting procedure). In particular, for the whole PLD graph ($6$M nodes), truncated power law is found strongly preferred over any other heavy tailed distribution. \citet{clauset2009power} analyzed the degree in the Web dataset from \citet{broder2000graph} and found the same result in that dataset as we find here: truncated power law was the best fit. Moreover, $x_{min}$ we find for indegree is $4$ and for outdegree $3$, meaning that in our case the fit describes a larger set of data points compared to less than 0.0001\% data points found to describe the power law in the distribution tail of \citet{meusel2015graph}. However, exactly because of the described differences, our results are not contradicting to those of \citet{meusel2015graph}. Namely, Meusel et al.\ have focused only on exploring the power law fits to their data (hence not investigating other heavy tailed distributions). Their reason is that indegree and outdegree distributions were explained by power law in earlier literature. For the same reason, they were concerned only with the tail of the distribution. Herein we present another type of insight: that a considerably larger portion of the data points in indegree and outdegree distributions can be better explained by another distribution, that is exponentially bounded power law. In other words, if one just wants to explore the power law, one must consider only the tail, but if we are concerned with explaining more of our data, then exponentially bounded power law is a better fit to indegree and outdegree.

The insights about the degree distributions in the whole PLD graph are relevant for the main focus of our analysis -- discerning the differences between the distributions of clean and malicious PLDs. As presented in bottom plots in Fig.\  \ref{fig:degree}, the truncated power law exponents are again different between those two classes. As with the number of pages, also now we find $\alpha_m < \alpha_c$ for both indegree and outdegree. Also, $\alpha_c > 2$ and $\alpha_m < 2$ indicating that the two distributions belong to different classes of power law-like distributions \citep{newman2005power}. Even if degree distributions are accurately known, this does not fully characterize the network \citep{WEB-017}. Still our insights call for further investigations on the differences between clean and malicious Web graph properties. The results for other network properties of PLDs are summarized in Table \ref{t:fit-netw}.

\begin{figure*}[!ht]
	\centering
	\begin{subfigure}{.49\linewidth}
		\centering
		\includegraphics[width=\linewidth]{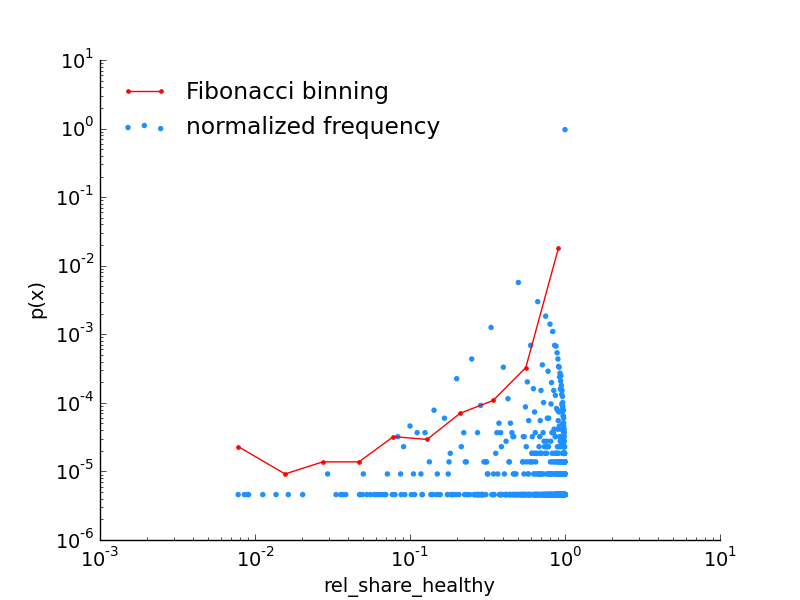}
	\end{subfigure} 
	\begin{subfigure}{.49\linewidth}
		\centering
		\includegraphics[width=\linewidth]{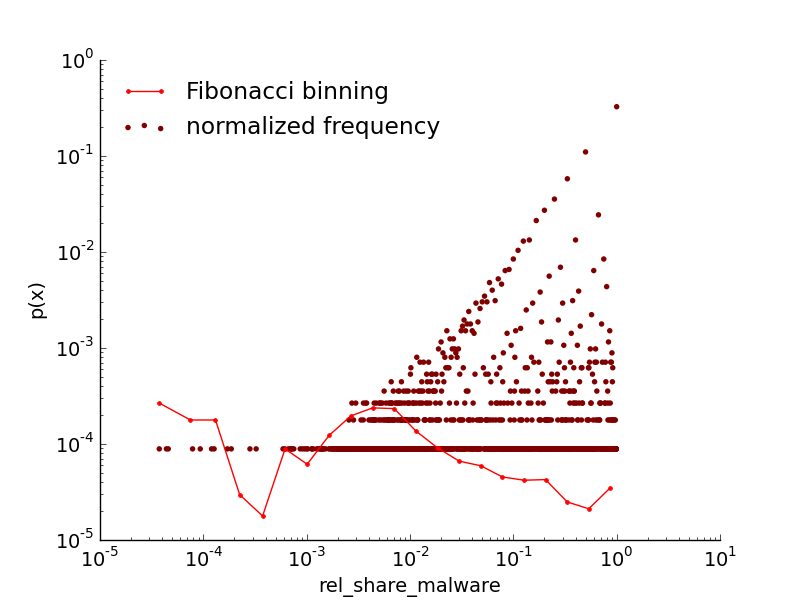}
	\end{subfigure}
	\caption{\textbf{PLD maliciousness: relative share of clean files on all PLDs (left) and relative share of malware on \textit{malicious} PLDs (right).} Distribution in the left plot reveals how majority of PLDs hosts mainly clean files. From the distribution in the right plot we notice a number of malicious PLDs devoted to serving almost only malware. However, Fibonacci binning \citep{vigna2013fibonacci} reveals that a majority of the malicious PLDs contains only around 1\% malicious files.}
	\label{fig:PLD_maliciousness}
\end{figure*}
\begin{table*}
	\caption{ \textbf{Statistics on top $5$ PLDs based on named properties in each subtable for:} total files number ($TF_{PLD}$ in Eq.\ \ref{eq:3}), malware files number (m), PageRank (PR), Alexa rank (Alexa) and relative maliciousness score ($\bar{r}$).
	}\label{t:extreme_PLDs} 
	\begin{tabular}{llllll} \hline
		\specialcell{\textbf{\texttt{PLD}}} & \textbf{\texttt{$TF_{PLD}$}} & \textbf{\texttt{m}} & 
		\textbf{\texttt{PR}} & \textbf{\texttt{Alexa}}& \textbf{\texttt{$\bar{r}$}} \\  \hline 
		\multicolumn{6}{c}{\texttt{PageRank}} \\  \hline
		\textbf{updatestar.com} & $4$ & $3$ & $1$&$8299$& $0.049$\\ 
		\textbf{facebook.com} & $3$ & $0$ & $2$ & $3$& $0.000$ \\ 
		\textbf{google.com} & $655$ & $85$ & $3$ & $1$& $0.004$ \\ 
		\textbf{googleapis.com } & $127$ & $5$ & $4$ & $1479$& $0.001$ \\ 
		\textbf{blogspot.com} & $1$ & $0$ & $5$ & $59$& $0.000$ \\ \hline
		\multicolumn{6}{c}{\texttt{Alexa rank}} \\  \hline
		\textbf{google.com} & $655$ & $85$ & $3$ & $1$& $0.004$ \\ 
		\textbf{youtube.com} & $2$ & $7$ & $8$ & $2$ & $0.000$\\
		\textbf{facebook.com} & $3$ & $0$ & $2$ & $3$& $0.000$ \\ 
		\textbf{baidu.com} & $103$ & $23$ & $21$ & $4$& $0.015$ \\ 
		\textbf{yahoo.com} & $10$ & $0$ & $17$ & $6$& $0.000$ \\ \hline
		\multicolumn{6}{c}{\texttt{$\bar{r}$}} \\  \hline
		\textbf{lao9123.com} & $2$ & $2$ & $>100K$ & $>1M$ & $0.910$ \\ 
		\textbf{188336.com} & $1$ & $1$ & $53,194$ & $>1M$ & $0.895$\\ 
		\textbf{chuangfa.cn} & $1$ & $1$ & $>100K$ & $>1M$& $0.891$ \\ 
		\textbf{starpoint.net} & $1$ & $1$ & $>100K$ & $>1M$& $0.859$ \\ 
		\textbf{tactearnhome.com} & $1$ & $1$ & $36\,824$ & $>1M$& $0.852$ \\ \hline
	\end{tabular}
	\begin{tabular}{llllll} \hline
		\specialcell{\textbf{\texttt{PLD}}} & \textbf{\texttt{$TF_{PLD}$}} & \textbf{\texttt{m}} & 
		\textbf{\texttt{PR}} & \textbf{\texttt{Alexa}}& {\textbf{\texttt{$\bar{r}$}}} \\  \hline 
		\multicolumn{6}{c}{\texttt{total files}} \\  \hline
		\textbf{youku.com} & $394\,083$ & $0$ & $52$&$174$& $0.000$\\ 
		\textbf{thelib.ru} & $874\,08$ & $4$ & $40\,790$ & $90\,811$& $0.000$ \\ 
		\textbf{royallib.com} & $68\,328$ & $3$ & $82\,265$ & $9\,185$& $0.000$ \\ 
		\textbf{xunzai.com} & $58\,681$ & $58\,224$ & $9\,794$ & $152\,137$& $0.282$ \\
		\textbf{maven.org} & $49\,635$ & $33$ & $985$ & $24\,344$& $0.000$ \\ \hline
		\multicolumn{6}{c}{\texttt{unique files}} \\  \hline
		\textbf{thelib.ru} & $874\,08$ & $4$ & $40\,790$ & $90\,811$& $0.000$ \\ 
		\textbf{royallib.com} & $68\,328$ & $3$ & $82\,265$ & $9\,185$& $0.000$ \\ 
		\textbf{maven.org} & $49\,635$ & $33$ & $985$ & $24\,344$& $0.000$ \\ 
		\textbf{java2s.com} & $25\,421$ & $2$ & $13\,158$ & $5\,816$& $0.000$ \\ 
		\textbf{3gpp.org} & $15\,732$ & $2$ & $1\,558$ & $45\,216$& $0.000$ \\ \hline
		\multicolumn{6}{c}{\texttt{malware files}} \\  \hline
		\textbf{xunzai.com} & $58\,681$ & $58\,224$ & $9\,794$ & $152\,137$& $0.282$ \\ 
		\textbf{crsky.com} & $20\,912$ & $7\,107$ & $1\,151$ & $7\,835$& $0.121$ \\
		\textbf{3234.com} & $9\,269$ & $2\,818$ & $3\,248$ & $107\,641$& $0.116$ \\ 
		\textbf{05sun.com} & $3\,983$ & $3\,949$ & $8\,736$ & $58\,694$& $0.546$ \\ 
		\textbf{cncrk.com} & $2\,850$ & $2\,818$ & $9\,192$ & $13\,751$& $0.652$ \\ \hline
	\end{tabular}
\end{table*}

\section{Relative PLD maliciousness}\label{sec:rel_mal_core}
Instead of strictly dividing PLDs to the clean and malicious ones, we can assign to each of them a relative maliciousness score $\bar{r}$ as introduced in Eq.\ \ref{eq:3}. To understand the need for such a relative score, we first look at relative share of clean and malicious files on a PLD in Fig.\ \ref{fig:PLD_maliciousness}. Relative share of clean files follows a unimodal distribution with a peak on right. Hence, we can see how not only a majority of PLDs are clean, but also most of them host a majority of clean files. Since attackers aim to spread their malware files to the otherwise regular domains, this result indicates that malicious PLDs in our previous dichotomization include many such compromised PLDs. Namely, PLDs that are devoted mainly to serving malware and created by attackers are likely to have only several files that are mainly malicious \citep{invernizzi2014nazca}. Relative share of malware, on the other hand, can only be measured on the malicious domains from our previous dichotomization. To this score we can also look as \textit{malware distribution rate} of a PLD. The malware distribution rate shown in Fig.\ \ref{fig:PLD_maliciousness} (right) is multimodal, with one peak on left and one in the middle. The peak in the middle likely corresponds to regular PLDs with many clean files that are infiltrated with a few malware files ($1\%$ score). The scatterplot of the distribution reveals, however, a number of PLDs with the score almost $100\%$ -- those are likely set up and maintained by attackers.

\begin{figure*}
	\centering
	\begin{subfigure}{.44\linewidth}
		\centering
		\includegraphics[width=\linewidth]{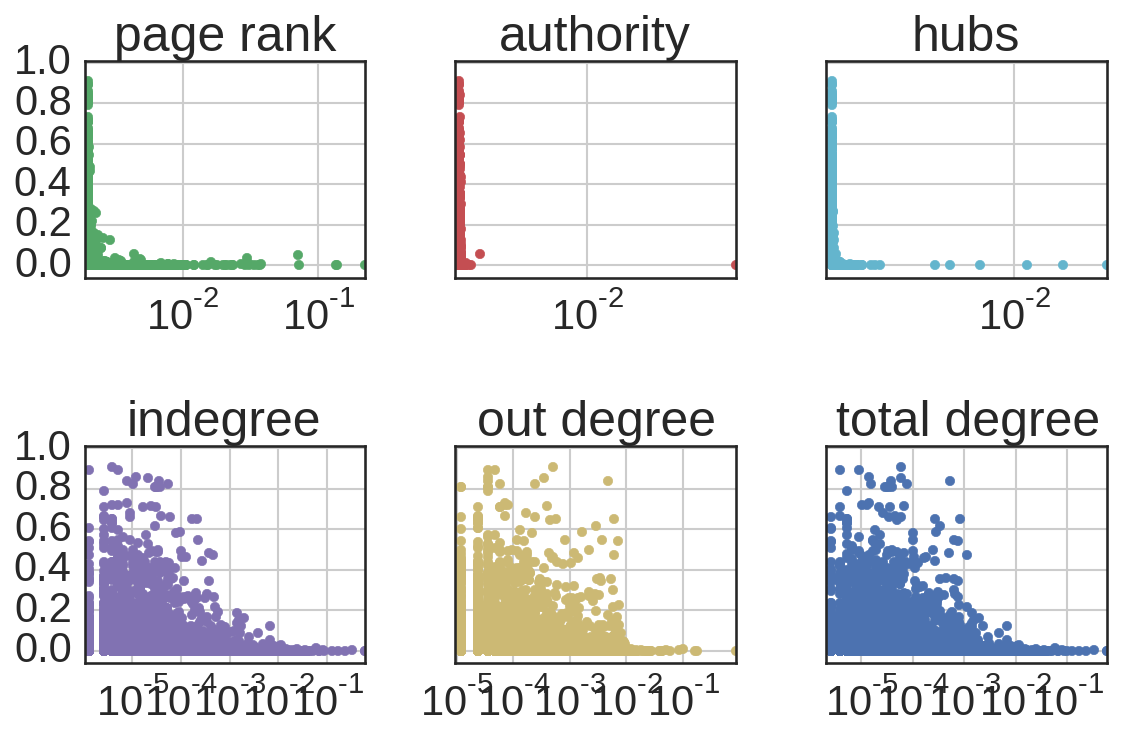}
	\end{subfigure}
	\begin{subfigure}{.44\linewidth}
		\centering
		\includegraphics[width=\linewidth]{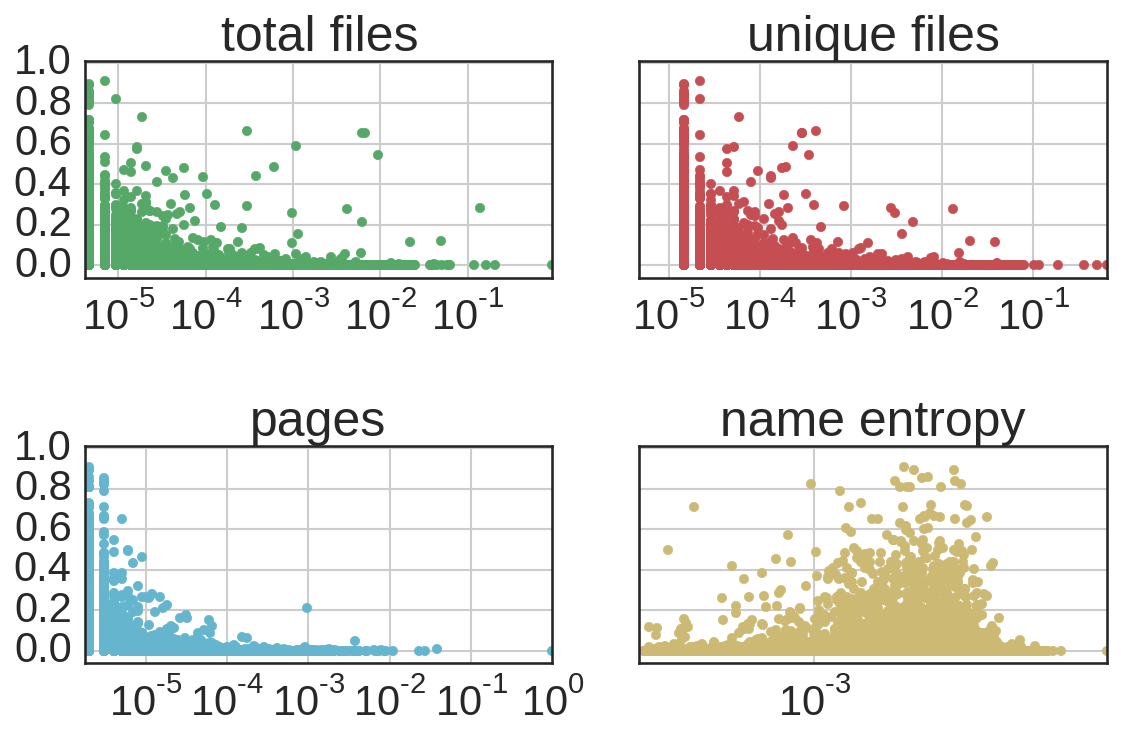}
	\end{subfigure}
	\caption{\textbf{Relationship between PLD (normalized) features and domain maliciousness score $\bar{r}$}. \label{fig:mal_score_all_fetures}}
\end{figure*}

Now we look at the relationship of previously analyzed features and this score. Fig.\ \ref{fig:mal_score_all_fetures} reveals that among most network central PLDs in $G$ there are no such with high $\bar{r}$. The observation holds for PLDs with the highest number of pages and total and unique files, too. The only of the assessed properties for which most malicious PLDs do not populate an extreme range is the name badness score. However, the most malicious domains are still found mainly within a particular range. In summary, these insights indicate a potential of the presented features in discerning the most malicious from clean PLDs.

Table \ref{t:extreme_PLDs} extends our insights into the maliciousness of PLDs. We find several malware files even on \texttt{google.com} and \texttt{baidu.com}, so they are marked as malicious in the strict scoring procedure. However, looking at their relative maliciousness score $\bar{r}$, that is low, we get a more accurate representation of their maliciousness. An interesting insight is also revealed from the PLDs with the highest number of malware files. Such PLDs host tens of thousands of malicious files, and still their scores $\bar{r}$ are not that high. This means that they host mostly suspicious and potentially unwanted files that are only marked by some AV engines, and not many highly malicious files.

\begin{figure*}
	\centering
	\includegraphics[width=0.79\textwidth]{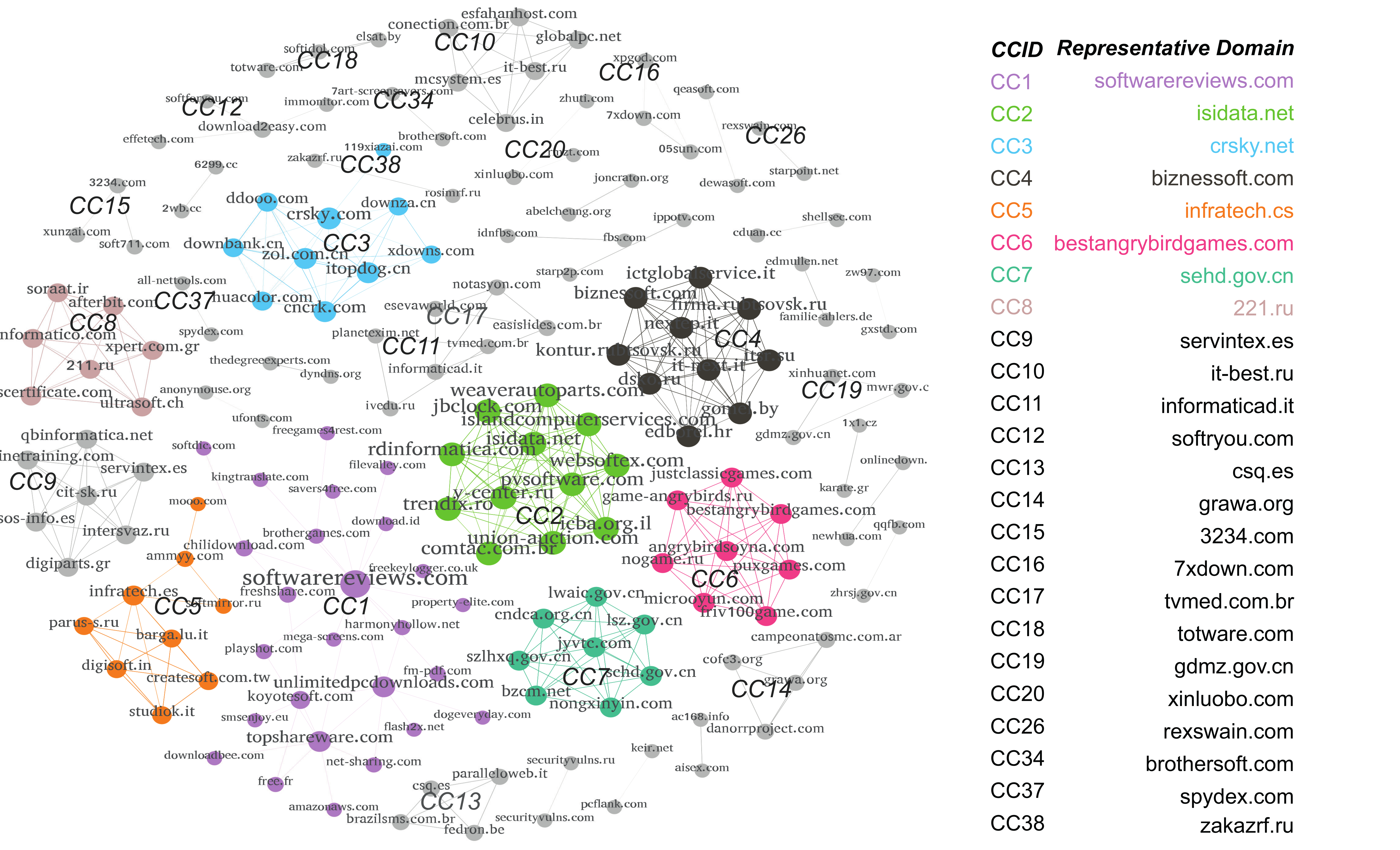}
	\caption{\textbf{Connected components (CC) in the \textit{domain malware co-occurrence network}.} Nodes in each \textbf{CC} share the same malware files. Node size is scaled with degree. In the legend, the \textbf{CC}s are in the decreasing order of size; up to the \textbf{CC8}, we use color code of the community; after that, only the \textbf{CCID}s are labeled on the graph. After \textbf{CC20}, we show only a few smaller components that the analysis revealed as interesting.}
	\label{fig:malware_domain_coocurrence}
\end{figure*}

\section{Malware co-occurrence network of PLDs} \label{sec:co-occur}
The \textit{malware co-occurrence network} $M$ reveals specific content delivery networks (CDNs) also sometimes called malware distribution networks (MDNs) \citep{zhang2011arrow}. They are used by attackers to manage a large number of malicious binaries, exploits and malware serving Web pages.

A visualization of the network $M$ is presented in Figure~\ref{fig:malware_domain_coocurrence}. $M$ consists of $40$ connected components: \textbf{CC1, ..., CC40} (in decreasing size as the index grows). The largest \textbf{CC1} has $26$ PLDs, while the \textbf{CCi}s for $i > 20$, have only a couple of PLDs. In addition to the score, each AV that is part of VirusTotal outputs its own textual description of the type of malware. For instance, consider the following output \textit{Trojan: Win32/Badur}, which hints that the malware file in question is a \textit{Trojan} of type \textit{Badur} that attacks the \textit{Windows} platform. By analyzing these textual outputs, we discover that each MDN is devoted to serving the type of malware with a particular purpose. 

\textbf{CC1} and \textbf{CC20} are serving mostly \textit{adwares, riskwares and undesirable software} that changes homepage, desktop background or search provider, such as {OpenCandy}, {Artemis}, {Somoto}, {Netcat}, and {Amonetize}.

In \textbf{CC2, CC3, CC4, CC5, CC8, CC9, CC10, CC13, CC16, CC17, CC18} and \textbf{CC19} besides potentially unwanted and adware files, we find more dangerous malware -- spywares and keyloggers. \textit{Spywares and keyloggers} can reveal passwords or even grant access to the user computer for a scammer, or they can cause browser redirection in an attempt to scam money from the victim, such as {Ammyy}, {Dafunk}, {Snoopit}, {Eldorado} and {Flystudio}. 

The domains in \textbf{CC6} and \textbf{CC7} employ \textit{drive-by download} approach in distributing malware. Namely, they serve malware that uses an Adobe Flash Player vulnerability to automatically download and run files once the victim visits their website.

\textit{Android malware} sharing was detected in \textbf{CC15}. The malware of type Trojan {Plankton} silently forwards information about the infected device to a remote location and when needed downloads additional files to the device. Another malware shared in this CC is {Ksappm}, suggested to be a Chinese based botnet used for malware distribution on Android devices\footnote{\url{http://androidmalwaredump.blogspot.fi/2013/01/androidtrojmdk-aka-androidksapp.html}}.

The domains in \textbf{CC34} share spyware for \textit{Mac OS X}, called {OpinionSpy}, this malware when installed on Mac could leak data and open a backdoor for further abuse.

\section{Predictive Power}
\label{sec:prediction_power}
Analyzing the set of all PLDs with an antivirus in the search for malware is computationally expensive. 
To reduce the number of PLDs to be analyzed, we estimate the probability of a PLD to contain malware based on its characteristics. 
This can be done by defining the task as a binary classification problem. The binary label represents if a PLD contains malware or not.
In this section, we present our results on predicting malware by describing how we label the domains, 
what is the evaluation metric, 
the different types of features used as input to the model 
and their performance.

The binary label for classification is defined as described in Section~\ref{p:dom_mal_score}. 
If a PLD has at least one file with a malicious score greater than zero, we mark it as a positive instance. Based on this rule, the dataset contains 5\% of PLDs labeled as malicious. We measure the performance of the models by measuring the Area Under the Curve (AUC) of the Receiver Operating Characteristic (ROC) function. 


As part of our experiments, we tried different sets of features, pre-processing steps (i.e., handling class imbalance and normalization) and classification algorithms.
The different sets of features are presented in Table~\ref{tab:feature_sets}.
To alleviate the class imbalance, 
we experiment with oversampling using SMOTE and random majority undersampling with replacement. 
In our predictions results, we present the performance of Gradient Boosting Trees. Other models (i.e., SVM, Logistic Regression) had lower performance. 

Following the ideas of \citep{castillo2007know}, we added stacked learning. The intuition behind stacked learning is that malicious PLDs are connected. Stacked learning has three steps. In the first step, we run the prediction model on the training set. In the second step, we calculate the average probability of each of the PLD neighbors. In the third step, we include the neighbors average probability as a new feature to the model and run the prediction model. Stacked learning does not leak information since everything is computed in the training dataset.

The data split consist of three parts. The training set contains 70\% the positive labels, the testing set contains 30\% the positive labels. A validation set is created using 30\% of the training set to fine tune the parameters of the Gradient Boosting Trees. 

\begin{table}
	\centering
	\caption{Feature sets used in the classification experiments}\label{tab:feature_sets}
	\begin{tabular}{p{0.2\linewidth}p{0.6\linewidth}} \hline
		\textbf{\texttt{Name}}&\textbf{\texttt{Features}}\\ \hline 
		\textbf{Centrality} & Authority, hubs, PageRank \\ 
		\textbf{Domain} &  Total files, unique files, num. pages, file distribution entropy, name entropy \\ 
		\textbf{Graph} &  Total degree, in degree, out degree, triangle count \\ 
		\textbf{Alexa Rank} &  Alexa rank, PLD in Alexa rank@1M (binary) \\ 
		\textbf{All} & All the features \\ \hline
	\end{tabular}
\end{table}

In the data pre-processing step we used imbalanced-learn \citep{imbalanced-learn}. For the classification models, we used Turi's GraphLab Create\footnote{\url{https://turi.com/}}. We also used GraphLab built-in functions for parameter tuning. The results of the experiments are presented in Table~\ref{tab:predictions_results}. For the best model (i.e., All + Stacked Learning), the feature importance of the model is presented in Table~\ref{tab:feature_importance}. By itself, the performance of our best model is modest. However, it can support a part of the detection process by reducing the suspicious PLD set without needing to parse HTML to extract content features (e.g., bag of words).


\begin{table*}
	\centering
	\begin{tabular}{llrllllrrrrr} \hline
		\textbf{\texttt{Features}} & \textbf{\texttt{Preprocessing}} & \textbf{\texttt{AUC}} & \textbf{\texttt{TP}} & \textbf{\texttt{TN}} & \textbf{\texttt{FP}} & \textbf{\texttt{FN}} & \textbf{\texttt{F1 Score}} & \textbf{\texttt{FNR}} & \textbf{\texttt{FPR}} & \textbf{\texttt{TNR}} & \textbf{\texttt{TPR}} \\ \hline 
		All + Stacked Learning &                      No preprocessing &  0.78 &          2\,304 &         54\,657 &          14\,751 &           	1\,318 &      0.22 &  0.36 &  0.21 &  0.79 &  0.64 \\ 
		All &                      No preprocessing &  0.76 &          2\,178 &         55\,324 &          14\,084 &           1\,444 &      0.22 &  0.40 &  0.20 &  0.80 &  0.60 \\ 
		Domain &                      No preprocessing &  0.74 &          2\,178 &         54\,694 &          14\,714 &           1\,444 &      0.21 &  0.40 &  0.21 &  0.79 &  0.60 \\ 
		Graph &                      No preprocessing &  0.65 &          1\,966 &         46\,099 &          23\,309 &           1\,656 &      0.14 &  0.46 &  0.34 &  0.66 &  0.54 \\ 
		Alexa Rank &  Undersampling* &  0.61 &          1\,514 &         52\,697 &          16\,711 &           2\,108 &      0.14 &  0.58 &  0.24 &  0.76 &  0.42 \\ 
		Centrality &                Undersampling &  0.60 &          1\,462 &         51\,966 &          17\,442 &           2\,160 &      0.13 &  0.60 &  0.25 &  0.75 &  0.40 \\ \hline
	\end{tabular}
	\caption{The best experiments for the feature set. For Alexa Rank an additional step of normalization was included.}
	\label{tab:predictions_results}
\end{table*}

\begin{table}
	\centering
	\begin{tabular}{p{0.7\linewidth}p{0.2\linewidth}}
		\hline
		\textbf{\texttt{Feature}} &  \textbf{\texttt{Count}} \\
		\hline 
		Name entropy &     83 \\ 
		PageRank &     71 \\ 
		Neighbors Probability (Stacked Learning) &     69 \\ 
		Indegree &     47 \\ 
		Alexa rank &     43 \\ 
		Outdegree &     41 \\ 
		Triangle count &     38 \\ 
		File distribution entropy &     34 \\ 
		Total files &     30 \\ 
		Total degree &     29 \\ 
		Unique files &     22 \\ 
		Authority &     21 \\ 
		Num. pages &     12 \\ 
		Hubs &      8 \\ 
		In Alex Rank (binary) &      0 \\ \hline
	\end{tabular}
	\caption{The feature's importance for the best model (i.e., All + Stacked Learning). The column \textit{count} is the sum of occurrence of the feature as a branching node in all trees.}
	\label{tab:feature_importance}
\end{table}

\section{Summary of results}
\label{sec:results}
Building upon the discussion of Sections~\ref{sec:rel_mal_core}, \ref{sec:co-occur} and \ref{sec:prediction_power}, we answer the research questions we set in the introduction.
\subsection{RQ1: Which theoretical distributions provide best fit for empirical distributions of local and network PLD features?}
\noindent
\textbf{Observation 1:} \textit{The number of pages, the number of unique files, indegree, outdegree and PageRank distributions are best explained by exponentially bounded power law (truncated power law) }.
While most of the earlier studies discussed and assessed a fit of, in particular, degree distributions to a power law, we find that in our Web crawl, a majority of the data points is better explained by exponentially bounded power law.

\subsection{RQ2: How are the characteristics and network properties different between clean and malicious PLDs?}
\noindent
\textbf{Observation 2:} \textit{In the case of all distributions following truncated power law, the exponent $\alpha_m$ of the malicious class is lower compared to the exponent $\alpha_c$ of the clean class }.
Moreover, for indegree, outdegree and the number of unique files, while $\alpha_c > 2$, at the same time $\alpha_m<2$, indicating a qualitatively different power law distribution.
\smallskip
\noindent
\smallskip

\textbf{Observation 3:} \textit{Maliciousness vs.\ centrality}.
The most malicious PLDs (i.e., those likely maintained by the attackers) do not have high values for network centrality nor local characteristics. However, attackers do manage to spread their malicious files to some of the most important and central regular PLDs.

\subsection{RQ3: What is the predictive power of PLD features?}
\noindent
\textbf{Observation 4:} \textit{Features Importance}.
By experimenting with different sets of features we obtaine their individual classification performance. Domain features (i.e., total files, unique files, num. pages, file distribution entropy, name entropy) are the best set of individual features, followed by graph features (i.e., total degree, in degree, out degree, triangle count), Alexa rank features (i.e., Alexa rank, PLD in Alexa rank@1M (binary)) and centrality features (i.e., authority, hubs, PageRank). 

When all the features are combined, the features importance is in the following order: name entropy, PageRank, neighbors probability (i.e., stacked learning), indegree, Alexa rank, outdegree, triangle count, file distribution entropy, total files, total degree, unique files, authority, num.\ pages and hubs.
\smallskip
\noindent
\smallskip

\textbf{Observation 5:} \textit{Model Performance}.
The best model in our experiment was Gradient Boosting Trees using all the features (i.e., centrality, domain, graph, Alexa rank) and a stacked learning step. The model achieved an \textit{AUC} of \textit{.78}. This model could be used as part of the detection process to reduce the number of suspicious PLD set to be fully analyzed by an antivirus. However, the model could not classify all of the PLDs with these features. A possible way to increase the prediction power is to include content (i.e., parsing the HTML) and adapting the crawling process \citep{invernizzi2012evilseed}.
\section{Discussion and conclusion}
\label{sec:discussion}
We presented results of data science application to a large Web crawl. Our results are a Web science contribution that increases the understanding on how different Web features are distributed -- we find that most of them are well explained by exponentially bounded power law. The size of the crawl and crawling policies might affect the observed distributions. Our data is smaller in size (around $6.5$ times) compared to those analyzed by \citet{meusel2015graph}. Also the crawling limitation of $1\,000$ hyperlinks from a website is particularly visible in the outdegree distribution. However, we still think that our insights in the crawl of this size are relevant for other researchers who might deal with datasets of similar size and possible limitations. Finally, power law degree distributions are shown to be a result of preferential attachment process during the graph growth \citep{barabasi1999emergence}, while power law degree distribution with an exponential cutoff results from competition-induced preferential attachment \citep{berger2005degree}. As discussed by \citet{d2007emergence} competition-induced preferential attachment better explains several real world degree distributions, including the Internet at the AS-level. Our results add Web degree distributions to that group.

We also show the difference in the exponents of the distributions pertaining to malicious versus clean websites, where malicious power law exponent is always lower. This result is a contribution to Web security as such knowledge can support the design of domain reputation classifiers and antivirus engines. In particular, we show such that even such content-agnostic features have discriminating power as features for machine learning prediction by achieving a relatively high AUC of 0.78. As future work, we plan to use temporal Web datasets in order to describe evolution of malicious activities and consequently offer more advanced recommendations for improving cyber security methods on the Web. Another line of future work is to add content features and apply targeted crawling to improve the malware classification performance.  

\subsection{Limitations}
Even if several Web distributions are shown to follow a power law, Web may not be scalefree in the sense that a sample crawl
is representative of the true Web degree distribution \citep{WEB-017}. We acknowledge that the crawl used in our study is limiting in that sense. In particular, crawling and sampling procedures induce biases \citep{achlioptas2009bias}, and we have not attempted to correct for those. Another crawling process limitation is that cloaking \citep{wang2011cloak} was not considered and that might limit our visibility to the malware files and websites. When it comes to the definition of what constitutes maliciousness of files and websites, we faced couple of additional trade-offs that should be pointed out. First is that only binary files of certain format and smaller size than a given threshold are downloaded. Hence, potential malware threats of other file type and size are not included in our definition. As discussed in the text, we applied the most strict definition for PLD maliciousness, which under the availability of a larger malware dataset should be tested in relaxed forms. 
\end{document}